\documentclass[conference,10pt,letterpaper]{IEEEtran}
\usepackage[english]{babel}
\usepackage[T1]{fontenc}
\usepackage{cite,url,color} 
\usepackage{graphics,amsfonts}
\usepackage{epstopdf}
\usepackage[pdftex]{graphicx}
\usepackage[cmex10]{amsmath}
\usepackage[utf8]{inputenc}
\usepackage{enumitem}

\usepackage{array}
\usepackage[font=scriptsize]{caption}
\usepackage[font=scriptsize]{subcaption}

\usepackage[top=1.5cm, bottom=2cm, right=1.6cm,left=1.6cm]{geometry}
\usepackage{indentfirst}

\usepackage{times}

\usepackage{breqn}
\usepackage{booktabs}
\usepackage{etoolbox}
\usepackage[acronyms,nonumberlist,nopostdot,nomain,nogroupskip]{glossaries}
\newacronym{kpi}{KPI}{Key Performance Indicator}
\newacronym{3gpp}{3GPP}{3rd Generation Partnership Project}
\newacronym{adc}{ADC}{Analog to Digital Converter}
\newacronym{5g}{5G}{5th generation}
\newacronym{aimd}{AIMD}{Additive Increase Multiplicative Decrease}
\newacronym{am}{AM}{Acknowledged Mode}
\newacronym{amc}{AMC}{Adaptive Modulation and Coding}
\newacronym{aqm}{AQM}{Active Queue Management}
\newacronym{awgn}{AGWN}{Additive White Gaussian Noise}
\newacronym{balia}{BALIA}{Balanced Link Adaptation}
\newacronym{bdp}{BDP}{Bandwidth-Delay Product}
\newacronym{bf}{BF}{Beamforming}
\newacronym{cc}{CC}{Congestion Control}
\newacronym{cdf}{CDF}{Cumulative Distribution Function}
\newacronym{cn}{CN}{Core Network}
\newacronym{cqi}{CQI}{Channel Quality Information}
\newacronym{cp}{CP}{Control Plane}
\newacronym{csirs}{CSI-RS}{Channel State Information - Reference Signal}
\newacronym{dc}{DC}{Dual Connectivity}
\newacronym{dce}{DCE}{Direct Code Execution}
\newacronym{dci}{DCI}{Downlink Control Information}
\newacronym{dl}{DL}{Downlink}
\newacronym{dmr}{DMR}{Deadline Miss Ratio}
\newacronym{dmrs}{DMRS}{DeModulation Reference Signal}
\newacronym{e2e}{E2E}{End-to-End}
\newacronym{ecn}{ECN}{Explicit Congestion Notification}
\newacronym{edf}{EDF}{Earliest Deadline First}
\newacronym{enb}{eNB}{evolved Node Base}
\newacronym{epc}{EPC}{Evolved Packet Core}
\newacronym{es}{ES}{Edge Server}
\newacronym{fdma}{FDMA}{Frequency Division Multiple Access}
\newacronym{fdd}{FDD}{Frequency Division Duplexing}
\newacronym[firstplural=Radio Access Technologies (RATs)]{rat}{RAT}{Radio Access Technology}
\newacronym{fs}{FS}{Fast Switching}
\newacronym{ftp}{FTP}{File Transfer Protocol}
\newacronym{gnb}{gNB}{Next Generation Node Base}
\newacronym{harq}{HARQ}{Hybrid Automatic Repeat reQuest}
\newacronym{hetnet}{HetNet}{Heterogeneous Network}
\newacronym{hh}{HH}{Hard Handover}
\newacronym{hol}{HOL}{Head-of-Line}
\newacronym{ia}{IA}{Initial Access}
\newacronym{imt}{IMT}{International Mobile Telecommunication}
\newacronym{iot}{IoT}{Internet of Things}
\newacronym{los}{LOS}{Line of Sight}
\newacronym{lte}{LTE}{Long Term Evolution}
\newacronym{m2m}{M2M}{Machine to Machine}
\newacronym{mac}{MAC}{Medium Access Control}
\newacronym{mc}{MC}{Multi-Connectivity}
\newacronym{mcs}{MCS}{Modulation and Coding Scheme}
\newacronym{mec}{MEC}{Mobile Edge Cloud}
\newacronym{mi}{MI}{Mutual Information}
\newacronym{mimo}{MIMO}{Multiple Input, Multiple Output}
\newacronym{mmwave}{mmWave}{millimeter wave}
\newacronym{mptcp}{MPTCP}{Multipath TCP}
\newacronym{mr}{MR}{Maximum Rate}
\newacronym{mss}{MSS}{Maximum Segment Size}
\newacronym{mtd}{MTD}{Machine-Type Device}
\newacronym{mtu}{MTU}{Maximum Transmission Unit}
\newacronym{nfv}{NFV}{Network Function Virtualization}
\newacronym{nlos}{NLOS}{Non Line of Sight}
\newacronym{nr}{NR}{New Radio}
\newacronym{ofdm}{OFDM}{Orthogonal Frequency Division Multiplexing}
\newacronym{pdcch}{PDCCH}{Physical Downlonk Control Channel}
\newacronym{pdcp}{PDCP}{Packet Data Convergence Protocol}
\newacronym{pdsch}{PDSCH}{Physical Downlink Shared Channel}
\newacronym{pdu}{PDU}{Packet Data Unit}
\newacronym{pf}{PF}{Proportional Fair}
\newacronym{pgw}{PGW}{Packet Gateway}
\newacronym{phy}{PHY}{Physical}
\newacronym{pbch}{PBCH}{Physical Broadcast Channel}
\newacronym[plural=\gls{mme}s,firstplural=Mobility Management Entities (MMEs)]{mme}{MME}{Mobility Management Entity}
\newacronym{prb}{PRB}{Physical Resource Block}
\newacronym{pss}{PSS}{Primary Synchronization Signal}
\newacronym{pucch}{PUCCH}{Physical Uplink Control Channel}
\newacronym{pusch}{PUSCH}{Physical Uplink Shared Channel}
\newacronym{rach}{RACH}{Random Access Channel}
\newacronym{ran}{RAN}{Radio Access Network}
\newacronym{red}{RED}{Random Early Detection}
\newacronym{rf}{RF}{Radio Frequency}
\newacronym{rlc}{RLC}{Radio Link Control}
\newacronym{rlf}{RLF}{Radio Link Failure}
\newacronym{rrc}{RRC}{Radio Resource Control}
\newacronym{rrm}{RRM}{Radio Resource Management}
\newacronym{rr}{RR}{Round Robin}
\newacronym{rs}{RS}{Remote Server}
\newacronym{rsrp}{RSRP}{Reference Signal Received Power}
\newacronym{rss}{RSS}{Received Signal Strength}
\newacronym{rtt}{RTT}{Round Trip Time}
\newacronym{rw}{RW}{Receive Window}
\newacronym{rx}{RX}{Receiver}
\newacronym{sa}{SA}{standalone}
\newacronym{sack}{SACK}{Selective Acknowledgment}
\newacronym{sap}{SAP}{Service Access Point}
\newacronym{sch}{SCH}{Secondary Cell Handover}
\newacronym{scoot}{SCOOT}{Split Cycle Offset Optimization Technique}
\newacronym{sdma}{SDMA}{Spatial Division Multiple Access}
\newacronym{sinr}{SINR}{Signal to Interference plus Noise Ratio}
\newacronym{sm}{SM}{Saturation Mode}
\newacronym{snr}{SNR}{Signal to Noise Ratio}
\newacronym{son}{SON}{Self-Organizing Network}
\newacronym{ss}{SS}{Synchronization Signal}
\newacronym{srs}{SRS}{Sounding Reference Signal}
\newacronym{sss}{SSS}{Secondary Synchronization Signal}
\newacronym{tb}{TB}{Transport Block}
\newacronym{tcp}{TCP}{Transmission Control Protocol}
\newacronym{tdd}{TDD}{Time Division Duplexing}
\newacronym{tdma}{TDMA}{Time Division Multiple Access}
\newacronym{tfl}{TfL}{Transport for London}
\newacronym{tm}{TM}{Transparent Mode}
\newacronym{trp}{TRP}{Transmitter Receiver Pair}
\newacronym{tti}{TTI}{Transmission Time Interval}
\newacronym{ttt}{TTT}{Time-to-Trigger}
\newacronym{tx}{TX}{Transmitter}
\newacronym{ue}{UE}{User Equipment}
\newacronym{ul}{UL}{Uplink}
\newacronym{uml}{UML}{Unified Modeling Language}
\newacronym{um}{UM}{Unacknowledged Mode}
\newacronym{utc}{UTC}{Urban Traffic Control}
\newacronym{vm}{VM}{Virtual Machine}
\newacronym{rsrq}{RSRQ}{Reference Signal Received Quality}
\newacronym{rssi}{RSSI}{Received Signal Strength Indicator}
\newacronym{crs}{CRS}{Cell Reference Signal}
\newacronym{comp}{CoMP}{Coordinated Multi-Point}
\newacronym{cran}{C-RAN}{Cloud \acrlong{ran}}
\newacronym{ca}{CA}{Carrier Aggregation}
\newacronym{cco}{CC}{Carrier Component}
\newacronym{nsa}{NSA}{Non Stand Alone}
\newacronym{embb}{eMBB}{Enhanced Mobility Broadband}
\newacronym{bsr}{BSR}{Buffer Status Report}
\newacronym{srb}{SRB}{Service Radio Bearer}
\newacronym{scm}{SCM}{Spatial Channel Model}
\newacronym{upa}{UPA}{Uniform Planar Array}
\usepackage{tikz}
\usepackage{pgfplots}
\pgfplotsset{compat=newest} 
\pgfplotsset{plot coordinates/math parser=false} 
\newlength\fheight
\newlength\fwidth
\usetikzlibrary{plotmarks,patterns,decorations.pathreplacing,backgrounds,calc,arrows,arrows.meta,spy,matrix}
\usepgfplotslibrary{patchplots,groupplots}
\usepackage{tikzscale}

\tikzstyle{startstop} = [rectangle, rounded corners, minimum width=2cm, minimum height=0.5cm,text centered, draw=black]
\tikzstyle{io} = [trapezium, trapezium left angle=70, trapezium right angle=110, minimum width=3cm, minimum height=1cm, text centered, draw=black]
\tikzstyle{process} = [rectangle, minimum width=2cm, minimum height=0.5cm, text centered, draw=black, align=center]
\tikzstyle{decision} = [ellipse, minimum width=2cm, minimum height=1cm, text centered, draw=black]
\tikzstyle{arrow} = [thick,<->,>=stealth]
\tikzstyle{line} = [thick,>=stealth]
\tikzstyle{darrow} = [thick,<->,>=stealth,dashed]
\tikzstyle{sarrow} = [thick,->,>=stealth]
\tikzstyle{larrow} = [line width=0.1mm,dashdotted,<->,>=stealth]

\definecolor{SchoolColor}{RGB}{0.71, 0, 0.106}
\definecolor{chaptergrey}{rgb}{0.61, 0, 0.09} 
\definecolor{midgrey}{rgb}{0.4, 0.4, 0.4}
\definecolor{chaptergreen}{rgb}{0.09, 0.612, 0}
\definecolor{chapterpurple}{rgb}{0.522, 0, 0.612}
\definecolor{chapterlightgreen}{rgb}{0, 0.612, 0.522}

\makeatletter
\def\grd@save@target#1{%
  \def\grd@target{#1}}
\def\grd@save@start#1{%
  \def\grd@start{#1}}
\tikzset{
  grid with coordinates/.style={
    to path={%
      \pgfextra{%
        \edef\grd@@target{(\tikztotarget)}%
        \tikz@scan@one@point\grd@save@target\grd@@target\relax
        \edef\grd@@start{(\tikztostart)}%
        \tikz@scan@one@point\grd@save@start\grd@@start\relax
        \draw[minor help lines] (\tikztostart) grid (\tikztotarget);
        \draw[major help lines] (\tikztostart) grid (\tikztotarget);
        \grd@start
        \pgfmathsetmacro{\grd@xa}{\the\pgf@x/1cm}
        \pgfmathsetmacro{\grd@ya}{\the\pgf@y/1cm}
        \grd@target
        \pgfmathsetmacro{\grd@xb}{\the\pgf@x/1cm}
        \pgfmathsetmacro{\grd@yb}{\the\pgf@y/1cm}
        \pgfmathsetmacro{\grd@xc}{\grd@xa + \pgfkeysvalueof{/tikz/grid with coordinates/major step x}}
        \pgfmathsetmacro{\grd@yc}{\grd@ya + \pgfkeysvalueof{/tikz/grid with coordinates/major step y}}
        \foreach \x in {\grd@xa,\grd@xc,...,\grd@xb}
        \node[anchor=north] at (\x,\grd@ya) {\pgfmathprintnumber{\x}};
        \foreach \y in {\grd@ya,\grd@yc,...,\grd@yb}
        \node[anchor=east] at (\grd@xa,\y) {\pgfmathprintnumber{\y}};
      }
    }
  },
  minor help lines/.style={
    help lines,
    gray,
    line cap =round,
    xstep=\pgfkeysvalueof{/tikz/grid with coordinates/minor step x},
    ystep=\pgfkeysvalueof{/tikz/grid with coordinates/minor step y}
  },
  major help lines/.style={
    help lines,
    line cap =round,
    line width=\pgfkeysvalueof{/tikz/grid with coordinates/major line width},
    xstep=\pgfkeysvalueof{/tikz/grid with coordinates/major step x},
    ystep=\pgfkeysvalueof{/tikz/grid with coordinates/major step y}
  },
  grid with coordinates/.cd,
  minor step x/.initial=.5,
  minor step y/.initial=.2,
  major step x/.initial=1,
  major step y/.initial=1,
  major line width/.initial=1pt,
}
\makeatother

\usepackage{mathtools}

\IEEEoverridecommandlockouts

\usepackage{placeins}
\newif\iftikz
\tikztrue
\begin{document}
\title{Impact of Channel Models on the End-to-End Performance of mmWave Cellular Networks}

\author{Michele Polese, Michele Zorzi\\
 \small Department of Information Engineering, University of Padova, Italy \\
e-mail: \{polesemi, zorzi\}@dei.unipd.it\vspace{-.4cm}
\thanks{This work was partially supported by the U.S. Department of Commerce/NIST (Award No. 70NANB17H166) and by the CloudVeneto initiative.}}


\maketitle

\glsunset{nr}

\tikzstyle{startstop} = [rectangle, rounded corners, minimum width=2cm, minimum height=0.5cm,text centered, draw=black]
\tikzstyle{io} = [trapezium, trapezium left angle=70, trapezium right angle=110, minimum width=3cm, minimum height=1cm, text centered, draw=black]
\tikzstyle{process} = [rectangle, minimum width=2cm, minimum height=0.5cm, text centered, draw=black, alignb=center]
\tikzstyle{decision} = [ellipse, minimum width=2cm, minimum height=1cm, text centered, draw=black]
\tikzstyle{arrow} = [thick,<->,>=stealth]
\tikzstyle{line} = [thick,>=stealth]
\tikzstyle{darrow} = [thick,<->,>=stealth,dashed]
\tikzstyle{sarrow} = [thick,->,>=stealth]
\tikzstyle{larrow} = [line width=0.05mm,dashdotted,>=stealth]

\begin{abstract}
Communication at mmWave frequencies is one of the major innovations of the fifth generation of cellular networks, because of the potential multi-gigabit data rate given by the large amounts of available bandwidth. The mmWave channel, however, makes reliable communications particularly challenging, given the harsh propagation environment and the sensitivity to blockage. Therefore, proper modeling of the mmWave channel is fundamental for accurate results in system simulations of mmWave cellular networks. Nonetheless, complex models, such as the 3GPP channel model for frequencies above 6 GHz, may introduce a significant overhead in terms of computational complexity. In this paper we investigate the trade offs related to the accuracy and the simplicity of the channel model in end-to-end network simulations, and the impact on the performance evaluation of transport protocols.
\end{abstract}


\begin{picture}(0,0)(0,-320)
\put(0,0){
\put(0,0){\footnotesize This paper has been accepted for publication at} 
\put(0,-10){\footnotesize IEEE SPAWC 2018, Kalamata, Greece, June 2018}}
\end{picture}

\section{Introduction}\label{sec:intro}
The next generation of cellular networks (5G) targets massive improvements in several \glspl{kpi} to sustain the mobile traffic growth and the new use cases that will emerge in the near future.
In particular, 5G networks will support very high throughput, combined with ultra-low latency, and high reliability and device density~\cite{osseiran2014scenarios,BocHLMP:14}. MmWave communications, together with the enhancements in spectral efficiency and network densification, are a possible enabler for the multi-gigabit-per-second data rates envisioned in future 5G networks~\cite{JerryPi:11}. Thanks to the wide availability of free spectrum at such high frequencies, network operators can allocate a much larger bandwidth with respect to the sub-6 GHz frequencies traditionally used for cellular communications.

The mmWave bands, however, are characterized by a harsh propagation environment that makes it difficult to reliably deploy a truly mobile mmWave network. The two main issues are the propagation loss, which is proportional to the square of the carrier frequency, and the blockage caused by common materials such as brick, mortar and also the human body~\cite{singh2007millimeter}. The first challenge can be addressed using directional antennas. Thanks to the smaller wavelengths, 
the same area can be packed with more antenna elements at mmWave frequencies than in the sub-6 GHz band, and, therefore, it is possible to increase the link budget with beamforming techniques~\cite{SunRap:cm14}. On the other hand, an ultra-dense deployment can help to avoid blockage phenomena and reduce the outage probability~\cite{AkdenizCapacity:14}.

New challenges related to the nature of the mmWave spectrum emerge also throughout the whole protocol stack. For example, the sudden transition between \gls{los} and \gls{nlos} states and the consequent drop in the channel quality may cause latency and efficiency issues to end-to-end data flows relying on the \gls{tcp}~\cite{zhang2016transport,polese2017tcp}, and the need for frequent and fast beam adaptation and/or handover calls for the design of efficient mobility management procedures~\cite{poleseHo}. Given the impact of mmWaves on the full stack, it is important to design and evaluate algorithms and protocols considering the performance of end-to-end systems. In this regard, network simulators are a valuable tool, whose reliability for wireless simulation, however, largely depends on the accuracy of the channel model~\cite{ferrand2016trends}.

In general, system level simulators consider a packet as a basic simulation unit, and do not model the actual bit transmission on the wireless link. The latter is usually abstracted with an error model, which maps the link \gls{sinr} on packet error probability curves to decide if the packet transmission was successful or not~\cite{brueninghaus2005link}. Therefore, the correct modeling of the transmission dynamics depends on the accuracy of the model for channel propagation and fading. 
In particular, given the characteristics of mmWave frequencies, the combined effect of propagation, fading and beamforming has a much higher impact on the end-to-end performance than in the sub-6 GHz band.

The publicly available ns-3 mmWave module~\cite{mezzavilla2018end}, that simulates 5G cellular networks at mmWave frequencies, features the implementation of the 3GPP channel model for frequencies above 6 GHz~\cite{38900,zhang20173gpp}. This is the channel model that, according to 3GPP, should be used for the simulations involving its latest standard, \gls{nr}, that also supports mmWave frequencies. However, it is a very complex model, which requires the generation of a large number of random numbers throughout the simulation, therefore limiting the scalability of the simulated scenarios. On the other hand, simpler channel models have been used in analytical studies in the literature, based on an abstraction of the beamforming gain and on Rayleigh or Nakagami fading~\cite{andrews2017modeling}. 

In this paper we compare the effects of the 3GPP channel model, which is used as a reference, and of a Nakagami fading-based channel model on the end-to-end performance of a mmWave cellular network. We show that there exists a trade off between the accuracy, defined as the difference in the considered metrics with respect to the reference 3GPP model, and the scalability of the simulations. In particular, the Nakagami-fading-based model introduces more severe fading than the 3GPP model, and, consequently, in the evaluated scenarios, always yields a lower throughput,
but it reduces the simulation execution time by at least an order of magnitude.

The remainder of the paper is organized as follows. In Sec.~\ref{sec:3gpp} we describe the main features of the 3GPP channel model, and review the models used in analytical studies. Then, in Sec.~\ref{sec:syst} we introduce the implementation of the channel model which will be compared against the 3GPP one, and present the results of the comparison in Sec.~\ref{sec:results}. Finally, in Sec.~\ref{sec:conclusions} we conclude the paper and address possible avenues of future work.

\section{Overview of mmWave Channel Models}\label{sec:3gpp}
In recent years, there have been several channel measurement campaigns at mmWave frequencies to characterize their propagation and fading, in different environments and conditions~\cite{Rappaport:13-BBmmW,weiler2014outdoor}. 
The main characteristics of the mmWave channel can be summarized as (i) a clear difference between \gls{los} and \gls{nlos} propagation; (ii) higher penetration loss than at sub-6 GHz frequencies; (iii) sparsity in the angular domain; and (iv) reduced impact of small scale fading~\cite{andrews2017modeling}. 
Several channel models, which capture the nature of mmWave propagation, have also been proposed, and
a review of the main contributions can be found in~\cite{hemadeh2018millimeter}. 

The modeling usually comprises a propagation loss model and a fading model. 
The propagation loss is computed by assigning to each physical location in the scenario a \gls{los} probability (unless deterministic environment models are used), and by applying different equations according to the \gls{los} or \gls{nlos} state of the user. A survey on path loss models can be found in~\cite{rappaport2017overview}. For fading, popular measurement-based channel models at mmWave frequencies~\cite{38900,metisChannel,jaeckel2014quadriga,AkdenizCapacity:14} are extensions of the WINNER and WINNER-II \glspl{scm}~\cite{winnerII}, while analytical studies generally use Rayleigh or Nakagami fading~\cite{bai2015coverage}. 

The 3GPP channel model for mmWave frequencies, which will be used as the reference model in this paper,
has been standardized in~\cite{38900}. This \gls{scm} is based on a channel matrix $\mathbf{H}$, whose entry $(i,j)$ represents the channel between the $i$-th and the $j$-th antenna elements at the transmitter and the receiver, respectively, and depends on the combined effect of $N$ multiple paths, i.e., the clusters. The clusters represent the direct \gls{los} path (if present) and the reflections that contribute to the total received power.  Each of them is modeled using a different delay and power, and is composed by multiple rays, distributed around a common cluster angle of arrival and departure. Moreover, two different components contribute to the modeling of fading~\cite{38900}: (i) large scale parameters, which are based on the user mobility and/or updates in the scenario, and affect the shadow fading, the delay spread of the clusters, the angular spread of the rays, and, in \gls{los}, the Ricean factor\footnote{The term \textit{Ricean factor} is used in~\cite{38900} to refer to the relative strength of the direct path with respect to the scattered components of a LOS channel, for any fading model (not necessarily Ricean).} associated to the \gls{los} cluster; and (ii) fast fading components, that model the small scale variations related to each cluster's delay and power, the actual angle of arrival and departure of the rays and the Doppler spread. 
The 3GPP channel model, and \glspl{scm} in general, can be easily integrated in the simulation with realistic beamforming, given that the beamforming gain can be computed by directly applying the beamforming vectors at the transmitter and receiver to the channel matrix $\mathbf{H}$~\cite{AkdenizCapacity:14}.

The parameters of the 3GPP channel are random variables generated from specific distributions, as detailed in~\cite{38900}. The total number of random variables drawn and, in general, of computations for each transmitter/receiver pair
in the scenario is proportional to $U\times S\times N$, with $U$ and $S$ the total number of antenna elements at the transmitter and the receiver. This can significantly increase the complexity of system level simulations, especially at mmWave frequencies, where antenna arrays with many elements are used to make up for the high propagation loss. 
In particular, random variables related to fast fading parameters are drawn at each transmission, while large scale fading parameters can be updated periodically\footnote{The report~\cite{38900} also introduces a spatially consistent procedure for the computation of the large scale fading parameters, which 
generates random numbers correlated with those drawn at the previous update based on the distance the user has covered between two consecutive updates.}.
Avoiding the update of the large scale fading parameters at each transmission is a practical assumption that helps decrease the computational complexity of the 3GPP channel model. However, despite the important efforts related to channel modeling at mmWave frequencies, there is a lack of statistical models of the rate at which large scale fading parameters evolve at mmWave frequencies, and only recently have results been presented in the case of specific blockage events~\cite{maccartney2017rapid}. Therefore, in the ns-3 implementation~\cite{zhang20173gpp}, it is possible to configure different large scale fading parameter update intervals.

Despite their accuracy and natural relation with beamforming techniques~\cite{ferrand2016trends}, \glspl{scm} cannot be used for analytical studies. Therefore, papers that investigate mmWave network performance analytically have proposed other channel abstractions, which use similar propagation loss equations, but simplify the modeling of fading and beamforming. For example, in~\cite{bai2015coverage}, the authors derive results on the coverage and rate of mmWave cellular networks using a channel model based on different path loss laws for \gls{los} and \gls{nlos}, Nakagami fading and a simplified sectored beamforming. Nakagami fading, introduced in~\cite{NAKAGAMI19603}, depends on a parameter $m$ which controls the amplitude of the fading phenomena: the larger $m$, the less severe the variations in amplitude.
The sectored beamforming model computes the beamforming gain $G$ by dividing the angular space in two regions: the main lobe, of angular width $\theta_b$ and maximum gain $G_M$, and the complementary sector with gain $G_m$. The total beamforming gain is given by the product of the transmitter and receiver gains, which can be either $G_M$ or $G_m$ according to the mutual position of the transmitter and the receiver. 
Similar approaches can be found in~\cite{singh2015tractable,andrews2017modeling,alkhateeb2017initial}. Other papers use Rayleigh fading for tractability, because it provides a lower bound to the system performance with respect to Nakagami fading in a stochastic geometry analysis~\cite{park2016tractable}. Nakagami fading, however, is generally preferred because it models more realistically the impact of fast fading on mmWave links~\cite{andrews2017modeling}, and returns Rayleigh fading for $m=1$. These channel models are usually computationally very efficient, because the number of random variables to be generated for each transmitter/receiver pair does not depend on the number of antenna elements and channel clusters.

To the best of our knowledge, \glspl{scm} and Nakagami-based models have not been compared for the evaluation of the performance of an end-to-end mmWave network. In~\cite{neil2017impact}, the authors compare different \glspl{scm} for mmWave cellular networks, but they only consider link level metrics for their evaluation. Similarly,~\cite{zeman2017accuracy} compares the different models available in the ns-3 mmWave module using as performance metric the \gls{mac} layer throughput of a single user. In this paper, we propose the implementation of a simple channel model, based on Nakagami fading, and compare it with the 3GPP channel model, focusing on the trade off between the accuracy of the simulation results and the computational complexity.

\section{System Model}\label{sec:syst}
The ns-3 mmWave module~\cite{mezzavilla2018end} can be used to simulate end-to-end mmWave networks, with realistic deployments, modeling of obstacles, a complete protocol stack for the \gls{ran}, a simple model for the core network and a full implementation of the \gls{tcp}/IP stack. The physical layer in the base stations and \glspl{ue} is based on \gls{ofdm}, and features a flexible frame structure~\cite{Dutta:15}, that can be adapted to simulate different 5G use cases~\cite{osseiran2014scenarios}, with a dynamic \gls{tdd} scheduling mechanism that adapts the duration of the scheduling intervals to the amount of data to be transmitted. It is also possible to simulate users' mobility and dual connectivity with \gls{lte} networks~\cite{poleseHo}. 

The module is equipped with different channel models~\cite{mezzavilla2018end}: there are two \glspl{scm}, i.e., the 3GPP~\cite{zhang20173gpp} and the NYU channel models~\cite{AkdenizCapacity:14}, and the possibility of using ray-tracing or measured channel traces. 
In addition to the available models, in this paper we introduce the \texttt{MmWaveSimpleChannel} class, which implements a channel model based on Nakagami fading and simplified beamforming, inspired to the widely-used models for mmWave cellular networks analysis in~\cite{singh2015tractable,alkhateeb2017initial,bai2015coverage,andrews2017modeling}. 

The \gls{sinr} for the link between the transmitter $i$ and the receiver $j$ is 
\begin{equation}
	P_{rx, j} = \frac{P_{tx, i} h_{i, j} G_{i, j} L_{i, j}}{\sigma^2 + \sum_{k \in \mathcal{I}} P_{tx, k} h_{k, j} G_{k, j} L_{k, j}},
\end{equation}
where $P_{tx, i}$ is the transmission power, $h_{i, j}$ the Nakagami fading, $G_{i,j}$ the beamforming gain and $L_{i, j}$ the pathloss for link $i, j$. The set $\mathcal{I}$ contains all the devices which are interfering with the transmission on link $i, j$, i.e., those that are actively transmitting during the same time interval and in the same frequency band as $i$ and $j$. 
Given that the modeling of the pathloss is not as computationally intensive as that of the fading, we use the 3GPP model for the propagation loss $L_{i,j}$, which depends on the \gls{los} condition, the 2D and 3D distance and the height of the devices, and the deployment scenario considered. Moreover, it is possible to select whether to enable or not the correlated shadowing~\cite{38900}.
Different $m$ values of the Nakagami fading for \gls{los} and \gls{nlos} links, respectively $m_{\rm LOS}$ and $m_{\rm NLOS}$, can be selected when configuring the simulation. We will provide insights on the choice of $m$ in the next section. 

Beamforming is modeled by computing the gain for \glspl{upa} with isotropic elements. In our future work, we plan to relax this modeling assumption and account for more realistic antenna patterns. Following the approach described in~\cite{37840,rebato2018study}, 
we compute the array radiation pattern $A_{A, i}(\theta^{i,j}, \phi^{i,j}, \theta^{i}_s, \phi^{i}_s)$ for device $i$ with respect to the signal to/from device $j$.
$\theta^{i,j}$ and $\phi^{i,j}$ are the vertical and horizontal angles associated to the \gls{los} direction between $i$ and $j$, and $\theta_{s}^{i}$ and $\phi_{s}^{i}$ are the vertical and horizontal steering angles for $i$. 
Since the antenna elements are isotropic, the array radiation pattern for an antenna array of $n$ elements is equal to the array factor (in dB)~\cite{rebato2018study}:
\begin{equation} 
\begin{split}
	A_{F,i}(\theta, \phi, \theta_s, \phi_s) &\;= 10 \log_{10} \left [1 + \rho(|\mathbf{a}(\theta, \phi)\mathbf{w}^T(\theta_s, \phi_s)|^2 - 1)  \right] \\ &\stackrel{\rho = 1}{=} 10 \log_{10} \left(|\mathbf{a}(\theta, \phi)\mathbf{w}^T(\theta_s, \phi_s)|^2 \right),
\end{split}
\end{equation}
where we consider $\rho = 1$ and omit the dependency on $i$ and $j$ in the angles. $\mathbf{a} \in \mathbb{C}^n$ represents the phase shift due to the placement of the antenna elements~\cite{37840}, i.e.,
\begin{equation}\begin{split}
	\mathbf{a}(\theta, \phi) = [a_{1,1}, a_{1,2}, \dots, a_{1, \sqrt{n}}, \dots, a_{\sqrt{n}, \sqrt{n}}], \;\mbox{where}\\ a_{p, r} = \frac{1}{\sqrt{n}}e^{j2\pi[(p-1)\cos(\theta)\Delta_v + (r-1)\sin(\theta)\sin(\phi)\Delta_h]},
\end{split}
\end{equation}
while $\mathbf{w} \in \mathbb{C}^n$ is the beamforming vector that weighs the antenna array elements to steer the beam in the desired direction, and is represented by
\begin{equation}\begin{split}
	\mathbf{w}(\theta_s, \phi_s) = [w_{1,1}, w_{1,2}, \dots, w_{1, \sqrt{n}}, \dots, w_{\sqrt{n}, \sqrt{n}}], \;\mbox{where}\\ w_{p, r} = e^{-j2\pi[(p-1)\cos(\theta_s)\Delta_v + (r-1)\sin(\theta_s)\sin(\phi_s)\Delta_h]}.
\end{split}
\end{equation}
The factors $\Delta_v$ and $\Delta_h$ are the vertical and horizontal antenna spacings normalized to the wavelength, and are both set to 0.5. We refer to~\cite{rebato2018study} for a discussion on the patterns that can be generated for different combinations of steering angle and directions.

In our implementation, we associate a beamforming vector to each endpoint of each pair of connected devices (i.e., a base station and the user connected to it), and we update it by setting the steering angles equal to those corresponding to the \gls{los} direction with a certain periodicity $T$, which is a parameter that can be set in the simulation scenario. By default, $T=20$~ms, one of the periodicities considered for the beamforming update in 3GPP \gls{nr}~\cite{moto2017ss}. Then, the beamforming gain $G_{i,j}$ is given (in dB) by $A_{A, i}(\theta^{i,j}, \phi^{i,j}, \theta^{i}_s, \phi^{i}_s) + A_{A, j}(\theta^{j,i}, \phi^{j,i}, \theta^{j}_s, \phi^{j}_s)$. The gain is maximum only if the devices are connected, i.e., if $i$ is transmitting to $j$ or vice versa, and the steering angles match the direction between $i$ and $j$. 

This beamforming model strikes a balance between complexity and flexibility. If the full channel matrix $\mathbf{H}$ is available in the simulation, as when using a \gls{scm}, then it is possible to model more realistically the beamforming gain, or, for example, compute the optimal beamforming vectors~\cite{zhang20173gpp}. However, the complexity involved with \gls{scm} is much higher, as we will discuss in the next section.

\section{Channel Model Comparison}\label{sec:results}

In this section, we consider two different simulation scenarios, with different end-to-end transport protocols. In the first, we deploy five base stations, in the center and at the four vertices of a square of side 200 m. $N_{\rm UE} \in \{2, 5, 10\}$ users are randomly placed in a disc around each base station, for a total of 10, 25 or 50 users. The base stations use a round robin scheduler. UDP is used as transport protocol to access data in a remote server, at a maximum rate of 400 Mbit/s per user. For the 3GPP channel model, the selected scenario is Urban Macro. The results are averaged over 20 independent runs, each with a simulated time of 10 s. 

\begin{figure}[t]
	\centering	
		\setlength{\fwidth}{.85\columnwidth}
		\setlength{\fheight}{0.45\columnwidth}
%
%
\definecolor{mycolor1}{rgb}{0.00000,0.44700,0.74100}%
\definecolor{mycolor4}{rgb}{0.6784,0.8471,0.9020}%
\definecolor{mycolor2}{rgb}{0.85000,0.32500,0.09800}%
\definecolor{mycolor5}{rgb}{1.0000,0.6275,0.4784}%
\definecolor{mycolor3}{rgb}{0.92900,0.69400,0.12500}%
\definecolor{mycolor6}{rgb}{1.0000,0.9804,0.8039}%

\begin{tikzpicture}
\pgfplotsset{every tick label/.append style={font=\scriptsize}}

\begin{axis}[%
width=0.951\fwidth,
height=\fheight,
at={(0\fwidth,0\fheight)},
scale only axis,
bar shift auto,
xmin=0.5,
xmax=3.5,
xtick=data,
xticklabels={10, 25, 50},
xlabel style={font=\scriptsize\color{white!15!black}},
xlabel={Number of users},
ymin=0,
ymax=0.00035,
ylabel style={font=\scriptsize\color{white!15!black}},
ylabel={Average MAC layer latency [s]},
axis background/.style={fill=white},
ylabel shift = -5 pt,
yticklabel shift = -2 pt,
xlabel shift = -4 pt,
xticklabel shift = -1 pt,
xmajorgrids,
ymajorgrids,
legend style={legend cell align=left, align=left, draw=white!15!black},
legend style={font=\scriptsize,legend cell align=left,align=left,draw=white!15!black, at={(0.01,0.01)},anchor=south west},
]
\addplot[ybar, bar width=0.178, fill=mycolor1, draw=black, fill opacity=0.2, area legend] table[row sep=crcr] {%
1	0.000332482324720709\\
2	0.000342044847221656\\
3	0.000343632969789475\\
};
\addplot[forget plot, color=white!15!black] table[row sep=crcr] {%
0.5	0\\
3.5	0\\
};

\addplot[ybar, bar width=0.178, fill=mycolor2, draw=black, fill opacity=0.2, area legend] table[row sep=crcr] {%
1	0.000332925518974359\\
2	0.000340313429845373\\
3	0.000343435327097597\\
};
\addplot[forget plot, color=white!15!black] table[row sep=crcr] {%
0.5	0\\
3.5	0\\
};

\addplot[ybar, bar width=0.178, fill=mycolor3, draw=black, fill opacity=0.2, area legend] table[row sep=crcr] {%
1	0.000331362877567104\\
2	0.000346241441816137\\
3	0.00034717592342534\\
};
\addplot[forget plot, color=white!15!black] table[row sep=crcr] {%
0.5	0\\
3.5	0\\
};

\addplot [color=black, draw=none, forget plot]
 plot [error bars/.cd, y dir = both, y explicit]
 table[row sep=crcr, y error plus index=2, y error minus index=3]{%
0.777777777777778	0.000332482324720709	5.38905632164467e-06	5.38905632164467e-06\\
1.77777777777778	0.000342044847221656	3.39488606721104e-06	3.39488606721104e-06\\
2.77777777777778	0.000343632969789475	1.30995795529792e-06	1.30995795529792e-06\\
};
\addplot [color=black, draw=none, forget plot]
 plot [error bars/.cd, y dir = both, y explicit]
 table[row sep=crcr, y error plus index=2, y error minus index=3]{%
1	0.000332925518974359	6.49463174970783e-06	6.49463174970783e-06\\
2	0.000340313429845373	5.26512996537926e-06	5.26512996537926e-06\\
3	0.000343435327097597	2.23484106425185e-06	2.23484106425185e-06\\
};
\addplot [color=black, draw=none, forget plot]
 plot [error bars/.cd, y dir = both, y explicit]
 table[row sep=crcr, y error plus index=2, y error minus index=3]{%
1.22222222222222	0.000331362877567104	5.81637675218057e-06	5.81637675218057e-06\\
2.22222222222222	0.000346241441816137	8.71789776298962e-07	8.71789776298962e-07\\
3.22222222222222	0.00034717592342534	1.66662939288935e-06	1.66662939288935e-06\\
};
\end{axis}

\begin{axis}[%
width=0.951\fwidth,
height=\fheight,
at={(0\fwidth,0\fheight)},
axis y line*=right,
axis x line=none,
hide x axis,
axis line style={-},
scale only axis,
bar shift auto,
xmin=0.5,
xmax=3.5,
xtick=data,
xticklabels={10, 25, 50},
xlabel style={font=\scriptsize\color{white!15!black}},
xlabel={Number of users},
ymin=0,
ymax=300000000,
ylabel style={font=\scriptsize\color{white!15!black}},
ylabel={User throughput [bit/s]},
ylabel shift = -5 pt,
yticklabel shift = -2 pt,
xlabel shift = -4 pt,
xticklabel shift = -1 pt,
legend style={legend cell align=left, align=left, draw=white!15!black},
legend style={font=\scriptsize,legend cell align=left,align=left,draw=white!15!black, at={(0.01,0.01)},anchor=south west},
]
\addplot [xshift=0.06cm, ybar, bar width=0.178, fill=mycolor1, draw=black, area legend, postaction={pattern=dots}] table[row sep=crcr] {%
1	235421815.099853\\
2	188716743.062277\\
3	133576859.256382\\
};
\addplot[forget plot, color=white!15!black] table[row sep=crcr] {%
0.5	0\\
3.5	0\\
};
\addlegendentry{Simple channel \sffamily{A}  -- $m_{\rm LOS} = 3$, $m_{\rm NLOS} = 2$}

\addplot [xshift=0.06cm, ybar, bar width=0.178, fill=mycolor2, draw=black, area legend, postaction={pattern=dots}] table[row sep=crcr] {%
1	256616196.529231\\
2	212660435.122503\\
3	142159549.912035\\
};
\addplot[forget plot, color=white!15!black] table[row sep=crcr] {%
0.5	0\\
3.5	0\\
};
\addlegendentry{Simple channel \sffamily{B}  -- $m_{\rm LOS} = 20$, $m_{\rm NLOS} = 10$}

\addplot [xshift=0.06cm, ybar, bar width=0.178, fill=mycolor3, draw=black, area legend, postaction={pattern=dots}] table[row sep=crcr] {%
1	267999715.439335\\
2	245174760.620487\\
3	159566733.505274\\
};
\addplot[forget plot, color=white!15!black] table[row sep=crcr] {%
0.5	0\\
3.5	0\\
};
\addlegendentry{3GPP channel}

\addplot [xshift=0.06cm, color=black, draw=none, forget plot]
 plot [error bars/.cd, y dir = both, y explicit]
 table[row sep=crcr, y error plus index=2, y error minus index=3]{%
0.777777777777778	235421815.099853	12134699.4360465	12134699.4360465\\
1.77777777777778	188716743.062277	7099484.29236688	7099484.29236688\\
2.77777777777778	133576859.256382	4555451.60399869	4555451.60399869\\
};
\addplot [xshift=0.06cm, color=black, draw=none, forget plot]
 plot [error bars/.cd, y dir = both, y explicit]
 table[row sep=crcr, y error plus index=2, y error minus index=3]{%
1	256616196.529231	12930509.2631813	12930509.2631813\\
2	212660435.122503	8068457.6892606	8068457.6892606\\
3	142159549.912035	5332862.78207906	5332862.78207906\\
};
\addplot [xshift=0.06cm, color=black, draw=none, forget plot]
 plot [error bars/.cd, y dir = both, y explicit]
 table[row sep=crcr, y error plus index=2, y error minus index=3]{%
1.22222222222222	267999715.439335	19478051.3728362	19478051.3728362\\
2.22222222222222	245174760.620487	11756655.7374642	11756655.7374642\\
3.22222222222222	159566733.505274	4908152.05221578	4908152.05221578\\
};
\end{axis}

\pgfplotsset{ticks=none}
\begin{axis}[%
ybar,
width=0.951\fwidth,
height=\fheight,
at={(0\fwidth,0\fheight)},
scale only axis,
xtick=data,
ymin=0,
ymax=600,
ylabel shift = -5 pt,
yticklabel shift = -2 pt,
legend style={font=\scriptsize,at={(0.01,0.35)},anchor=south west,legend cell align=left,align=left,draw=white!15!black},
enlarge x limits=0.15,
hide y axis,
hide x axis,
legend columns=2,
]
\addplot [fill=white,postaction={pattern=crosshatch dots}]
  table[row sep=crcr]{%
  1 0\\
  2 0\\
  3 0\\
  4 0\\
};
\addlegendentry{Throughput}

\addplot [fill=white]
  table[row sep=crcr]{%
  1 0\\
  2 0\\
  3 0\\
  4 0\\
};
\addlegendentry{Latency}
\end{axis}
\end{tikzpicture}%
		\caption{Average user throughput and latency for different channel models and number of users, in the scenario with UDP as transport.}
		\label{fig:thUdp}
\end{figure}
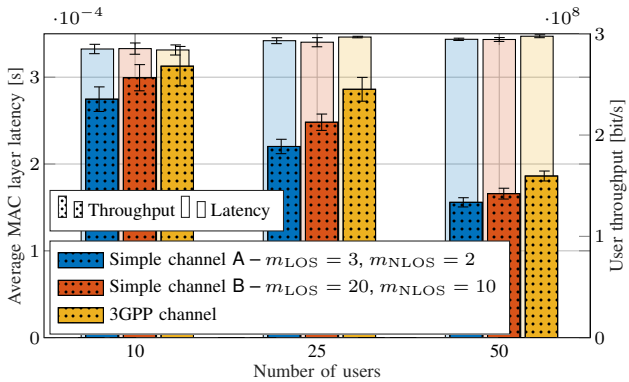

\begin{figure}
	\centering
	\setlength{\fwidth}{.85\columnwidth}
	\setlength{\fheight}{0.45\columnwidth}
	\input{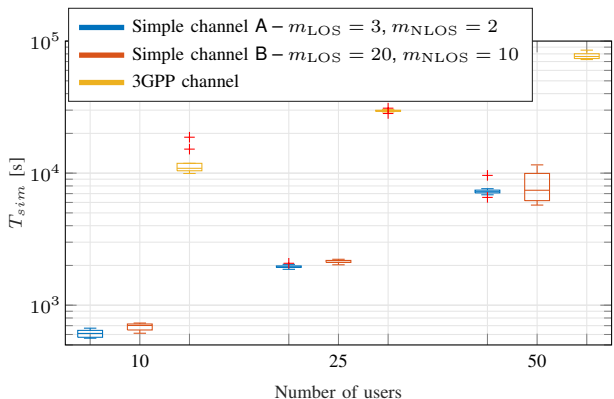}
	\caption{Boxplot for the simulation execution time, for different channel models and number of users, in the scenario with UDP as transport.}
	\label{fig:boxplot}
\end{figure}

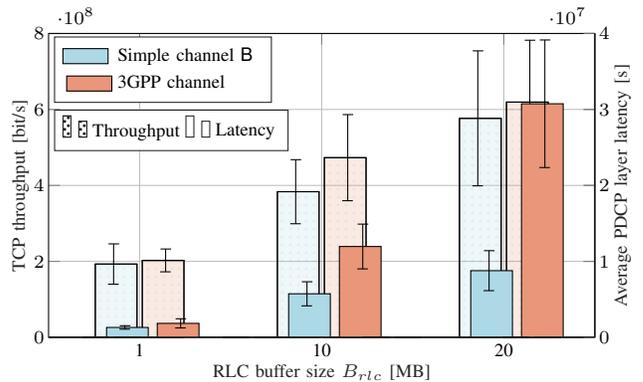
\begin{figure}[t]
	\centering	
		\setlength{\fwidth}{.85\columnwidth}
		\setlength{\fheight}{0.45\columnwidth}
%
%
\definecolor{mycolor1}{rgb}{0.6784,0.8471,0.9020}%
\definecolor{mycolor2}{rgb}{0.9137,0.5882,0.4784}%
\begin{tikzpicture}
\pgfplotsset{every tick label/.append style={font=\scriptsize}}

\begin{axis}[%
width=0.951\fwidth,
height=\fheight,
at={(0\fwidth,0\fheight)},
bar shift auto,
scale only axis,
xmin=0.5,
xmax=3.5,
xtick={1, 2, 3},
xticklabels={1, 10, 20},
xlabel={RLC buffer size $B_{rlc}$ [MB]},
xlabel style={font=\scriptsize\color{white!15!black}},
ymin=0,
ymax=800000000,
ylabel style={font=\color{white!15!black}},
ylabel style={font=\scriptsize\color{white!15!black}},
ylabel={TCP throughput [bit/s]},
axis background/.style={fill=white},
ylabel shift = -5 pt,
yticklabel shift = -2 pt,
xlabel shift = -4 pt,
xticklabel shift = -1 pt,
xmajorgrids,
ymajorgrids,
legend style={legend cell align=left, align=left, draw=white!15!black},
legend style={font=\scriptsize,legend cell align=left,align=left,draw=white!15!black, at={(0.01,0.01)},anchor=south west},
]
\addplot[ybar, bar width=0.229, fill=mycolor1, fill opacity=0.2, postaction={pattern=dots}, draw=black, area legend] table[row sep=crcr] {%
1	192992854.809054\\
2	383430131.015505\\
3	576576291.942543\\
};
\addplot[forget plot, color=white!15!black] table[row sep=crcr] {%
0.5	0\\
3.5	0\\
};

\addplot[ybar, bar width=0.229, fill=mycolor2, fill opacity=0.2, postaction={pattern=dots}, draw=black, area legend] table[row sep=crcr] {%
1	202581435.152617\\
2	473075338.096232\\
3	619389173.859948\\
};
\addplot[forget plot, color=white!15!black] table[row sep=crcr] {%
0.5	0\\
3.5	0\\
};

\addplot [color=black, draw=none, forget plot]
 plot [error bars/.cd, y dir = both, y explicit]
 table[row sep=crcr, y error plus index=2, y error minus index=3]{%
0.857142857142857	192992854.809054	53106441.9538195	53106441.9538195\\
1.85714285714286	383430131.015505	84351258.9132719	84351258.9132719\\
2.85714285714286	576576291.942543	177656973.322245	177656973.322245\\
};

\addplot [color=black, draw=none, forget plot]
 plot [error bars/.cd, y dir = both, y explicit]
 table[row sep=crcr, y error plus index=2, y error minus index=3]{%
1.14285714285714	202581435.152617	30038816.6368303	30038816.6368303\\
2.14285714285714	473075338.096232	113389030.083297	113389030.083297\\
3.14285714285714	619389173.859948	162804311.98503	162804311.98503\\
};

\end{axis}

\begin{axis}[%
width=0.951\fwidth,
height=\fheight,
at={(0\fwidth,0\fheight)},
axis y line*=right,
axis x line=none,
hide x axis,
axis line style={-},
scale only axis,
bar shift auto,
xmin=0.5,
xmax=3.5,
xtick={1, 2, 3},
xticklabels={1, 10, 20},
xlabel={RLC buffer size \sffamily{B}  [MB]},
xlabel style={font=\scriptsize\color{white!15!black}},
ylabel style={font=\scriptsize\color{white!15!black}},
ylabel={Average PDCP layer latency [s]},
ymin=0,
ymax=40000000,
ylabel shift = -5 pt,
yticklabel shift = -2 pt,
xlabel shift = -4 pt,
xticklabel shift = -1 pt,
legend style={font=\scriptsize,legend cell align=left,align=left,draw=white!15!black, at={(0.01,0.99)},anchor=north west},
]
\addplot[xshift=0.15cm, ybar, bar width=0.229, fill=mycolor1, draw=black, area legend] table[row sep=crcr] {%
1	1298622.21023434\\
2	5727796.12743126\\
3	8772489.93573632\\
};
\addplot[forget plot, color=white!15!black] table[row sep=crcr] {%
0.5	0\\
3.5	0\\
};
\addlegendentry{Simple channel \sffamily{B} }

\addplot[xshift=0.2cm, ybar, bar width=0.229, fill=mycolor2, draw=black, area legend] table[row sep=crcr] {%
1	1836033.97225996\\
2	11953409.9425159\\
3	30742662.7176681\\
};
\addplot[forget plot, color=white!15!black] table[row sep=crcr] {%
0.5	0\\
3.5	0\\
};
\addlegendentry{3GPP channel}

\addplot [xshift=0.15cm, color=black, draw=none, forget plot]
 plot [error bars/.cd, y dir = both, y explicit]
 table[row sep=crcr, y error plus index=2, y error minus index=3]{%
0.857142857142857	1298622.21023434	229612.406591024	229612.406591024\\
1.85714285714286	5727796.12743126	1587179.22179009	1587179.22179009\\
2.85714285714286	8772489.93573632	2631384.7556346	2631384.7556346\\
};

\addplot [xshift=0.2cm, color=black, draw=none, forget plot]
 plot [error bars/.cd, y dir = both, y explicit]
 table[row sep=crcr, y error plus index=2, y error minus index=3]{%
1.14285714285714	1836033.97225996	602739.669413098	602739.669413098\\
2.14285714285714	11953409.9425159	2948043.13963923	2948043.13963923\\
3.14285714285714	30742662.7176681	8396334.01796472	8396334.01796472\\
};
\end{axis}

\pgfplotsset{ticks=none}
\begin{axis}[%
ybar,
width=0.951\fwidth,
height=\fheight,
at={(0\fwidth,0\fheight)},
scale only axis,
xtick=data,
ymin=0,
ymax=600,
ylabel shift = -5 pt,
yticklabel shift = -2 pt,
legend style={font=\scriptsize,at={(0.01,0.75)},anchor=north west,legend cell align=left,align=left,draw=white!15!black},
enlarge x limits=0.15,
hide y axis,
hide x axis,
legend columns=2,
]
\addplot [fill=white,postaction={pattern=crosshatch dots}]
  table[row sep=crcr]{%
  1 0\\
  2 0\\
  3 0\\
  4 0\\
};
\addlegendentry{Throughput}

\addplot [fill=white]
  table[row sep=crcr]{%
  1 0\\
  2 0\\
  3 0\\
  4 0\\
};
\addlegendentry{Latency}
\end{axis}

\end{tikzpicture}%
		\caption{Average user throughput and latency for different channel models and \gls{rlc} buffer size $B_{rlc}$, in the scenario with TCP as transport.}
		\label{fig:thTcp}
\end{figure}

We compare the average user throughput and latency in Fig.~\ref{fig:thUdp} for different channel models. We test the 3GPP channel model and different $m$ values for the Nakagami fading in the simple channel model, i.e., setting {\sffamily{A}\sffamily} with $m_{\rm LOS} = 3$, $m_{\rm NLOS} = 2$~\cite{bai2015coverage} and setting {\sffamily{B}\sffamily} with $m_{\rm LOS} = 20$, $m_{\rm NLOS} = 10$~\cite{gupta2016on}, to simulate different impacts of the fading on the received signal. As it can be seen, there is no significant difference on the average \gls{mac} layer latency, which is in line with the results in~\cite{Dutta:15}. For the throughput, instead, the simple channel model always shows an average throughput smaller than that of the 3GPP model. Therefore, the Nakagami fading-based channel represents in the simulated scenarios a conservative bound with respect to the measurement-based 3GPP channel. In particular, it can be seen that the throughput decreases as the severity of the fading phenomena increases. For the simple channel the throughput loss ranges from 13\% (for 10 \glspl{ue}) to 23\% (for 25 \glspl{ue}) with setting {\sffamily{A}\sffamily}, and from 4\% (for 10 \glspl{ue}) to 13\% (for 25 \glspl{ue}) with the less conservative setting {\sffamily{B}\sffamily}. Therefore, the simple channel model described in Sec.~\ref{sec:syst} is less accurate than the reference 3GPP model, given that it introduces more severe fading, even though it exhibits the same trend when varying the number of users in the simulation. However, as shown in Fig.~\ref{fig:boxplot}, its usage reduces the simulation execution time $T_{sim}$ by an order of magnitude: for example, with 50 \glspl{ue}, $T_{sim}$ is 123 minutes and 33 seconds with the simple channel with configuration {\sffamily{A}\sffamily}, and 1183 minutes and 30 seconds with the 3GPP channel\footnote{All the simulations were run on a system with an Intel Xeon E5-2670v2 2.5 GHz CPU and 40 GB of RAM.}. 

The second scenario, instead, involves a single user, three mmWave and one LTE base stations deployed as in~\cite{tcp-mobicom}. The user moves in the scenario along a straight line for 100 m, and hands over between the different base stations. TCP NewReno is used as transport protocol, and the results for the throughput and latency with different \gls{rlc} buffer sizes $B_{rlc} \in \{1, 10, 20\}$~MB and channel models (the 3GPP and the simple channel model with setting {\sffamily{B}\sffamily}) are shown in Fig~\ref{fig:thTcp}. In general, the 3GPP channel model has higher throughput and latency with respect to the simple model, even though for a small buffer size $B_{rlc}=1$~MB the performance is similar. For both channel models latency and throughput increase with the buffer size, as expected~\cite{tcp-mobicom}, but the latency increase is higher with the 3GPP model. This is due to TCP's behavior during the slow start phase, in which the number of bytes that are sent doubles at each \gls{rtt}, until a packet is lost or the slow start threshold is reached. In these simulations, the latter is set to the maximum possible value of the TCP congestion window, thus TCP exits the slow start phase only because of packet loss. The 3GPP channel model incurs less severe fading than the simple channel model, therefore the probability of losing a packet because of random channel quality drops is smaller, and the slow start continues until the packet loss is triggered by an \gls{rlc} buffer overflow. The consequence is that there is a latency spike due to excessive buffering at the \gls{rlc} layer, which increases with the buffer size. On the other hand, with the simple channel model, TCP may exit earlier from slow start, and avoid overflowing the buffer and causing the initial latency spike, at the price of a reduced throughput.

\section{Conclusions}\label{sec:conclusions}
In this paper, we studied the impact of different channel models on end-to-end metrics and simulation complexity. In particular, we used the ns-3 mmWave module, and the already implemented 3GPP channel model as a reference, and proposed the implementation of a \textit{simpler} channel model, similar to those used for the mathematical analysis of mmWave networks, based on Nakagami fading and the computation of the beamforming gain from the array radiation pattern.

We showed that, with respect to the reference 3GPP model, the Nakagami fading-based model yields a lower throughput in end-to-end simulations with randomly generated scenarios, and both TCP and UDP as transport protocols. Moreover, when considering TCP, the different behavior and severity of the fading model generates different latency results, especially when larger buffer sizes in the base stations are considered. However, by using the simpler model it is possible to reduce the simulation execution time by an order of magnitude.

We believe that the insights we provided in this paper can provide guidance on which channel model should be used in mmWave simulations. When complex interactions between the channel and the end-to-end transport protocols are expected, it is better to use the reference 3GPP channel model, at the price of higher complexity. On the other hand, when the interplay between the transport and the lower layers is less complicated, a simpler model can be used to scale the simulations to a larger number of users, keeping in mind that the simulated throughput could be a lower bound with respect to that generated with the 3GPP channel. 

As future work, we plan to develop a statistical model that can better abstract the behavior of the 3GPP model while reducing the simulation complexity, and continue to investigate which are the main accuracy trade offs with respect to the 3GPP channel model. 
\bibliographystyle{IEEEtran}
\bibliography{IEEEabrv,bibl.bib}

\end{document}